\documentclass[pdflatex,sn-nature]{sn-jnl}% Style for submissions to Nature Portfolio journals
\usepackage{graphicx}%
\usepackage{multirow}%
\usepackage{amsmath,amssymb,amsfonts}%
\usepackage{amsthm}%
\usepackage{mathrsfs}%
\usepackage[title]{appendix}%
\usepackage{xcolor}%
\usepackage{textcomp}%
\usepackage{manyfoot}%
\usepackage{booktabs}%
\usepackage{algorithm}%
\usepackage{algorithmicx}%
\usepackage{algpseudocode}%
\usepackage{listings}%
\usepackage{pdfpages}%

\usepackage{geometry}
\pdfoutput=1
\theoremstyle{thmstyleone}%
\theoremstyle{thmstyletwo}%

\theoremstyle{thmstylethree}%

\begin{document}

\title[Article Title]{SA-GAT-SR: Self-Adaptable Graph Attention Networks  with Symbolic Regression for high-fidelity material property prediction}

%%=============================================================%%
%% GivenName	-> \fnm{Joergen W.}
%% Particle	-> \spfx{van der} -> surname prefix
%% FamilyName	-> \sur{Ploeg}
%% Suffix	-> \sfx{IV}
%% \author*[1,2]{\fnm{Joergen W.} \spfx{van der} \sur{Ploeg} 
%%  \sfx{IV}}\email{iauthor@gmail.com}
%%=============================================================%%

\author[1]{Junchi Liu}

\author[2]{Ying Tang}
\author*[3]{Sergei Tretiak}\email{Serg@lanl.gov}

\author*[4,5,6]{Wenhui Duan}\email{duanw@tsinghua.edu.cn}
\author*[1,2]{Liujiang Zhou}\email{ljzhou@uestc.edu.cn}

\affil[1]{School of Physics, State Key Laboratory of Electronic Thin Films and Integrated Devices, University of Electronic Science and Technology of China, Chengdu 611731, China}

\affil[2]{Institute of Fundamental and Frontier Sciences, University of Electronic Sciences and Technology of China, Chengdu, 611731, China}

\affil[3]{Theoretical Division, Center for Nonlinear Studies, and Center for Integrated Nanotechnologies, Los Alamos National Laboratory, Los Alamos, New Mexico 87545, United States}

\affil[4]{State Key Laboratory of Low Dimensional Quantum Physics and Department of Physics, Tsinghua University, Beijing 100084, China}

\affil[5]{Institute for Advanced Study, Tsinghua University, Beijing 100084, China}
\affil[6]{Frontier Science Center for Quantum Information, Beijing 100084, China}

%%==================================%%
%% Sample for unstructured abstract %%
%%==================================%%

\abstract{
    Recent advances in machine learning have demonstrated an enormous utility of deep learning approaches, particularly Graph Neural Networks (GNNs) for materials science. These methods have emerged as powerful tools for high-throughput prediction of material properties, offering a compelling enhancement and alternative to traditional first-principles calculations. While the community has predominantly focused on developing increasingly complex and universal models to enhance predictive accuracy, such approaches often lack physical interpretability and insights into materials behavior. Here, we introduce a novel computational paradigm—Self-Adaptable Graph Attention Networks integrated with Symbolic Regression (SA-GAT-SR)—that synergistically combines the predictive capability of GNNs with the interpretative power of symbolic regression. Our framework employs a self-adaptable encoding algorithm that automatically identifies and adjust attention weights so as to screen critical features from an expansive 180-dimensional feature space while maintaining $O(n)$ computational scaling. The integrated SR module subsequently distills these features into compact analytical expressions that explicitly reveal quantum-mechanically meaningful relationships, achieving 23$\times$ acceleration compared to conventional SR implementations that heavily rely on first principle calculations-derived features as input. This work suggests a new framework in computational materials science, bridging the gap between predictive accuracy and physical interpretability, offering valuable physical insights into material behavior.
}

\maketitle
\section*{Introduction}

The modern discovery of advanced functional materials demands predictive frameworks that successfully reconcile frequently conflicting requirements: quantum-level precision, generalizable physical insights, and human-interpretable design principles. While conventional computational methods struggle to address this challenge, machine learning (ML) approaches have shown remarkable progress across a wide range of materials property prediction tasks. Specifically, two distinct ML paradigms have emerged in solid state materials research: symbolic regression (SR) offering equation-based interpretability and graph neural networks (GNNs) excelling in structure-property mapping accuracy. SR reveals fundamental correlations through transparent mathematical descriptors to generate explicit descriptors. These descriptors are expressed as mathematical relationships among various material features, providing actionable insights for rational material design to achieve desired properties. However, the explicit feature combination in SR limits its ability to capture complex relationships among features in high-dimensional spaces. In contrast, GNNs circumvent this limitation through automated learning of hidden material representations via message-passing architectures\cite{bib20,bib21,bib24,bib26}, but their black-box nature obscures the underpinning atomic-scale physics driving property variations—a critical barrier for scientific discovery. This inherent trade-off between interpretative clarity (SR) and automated learning capability (GNN) has constrained computational materials science to situation-specific solutions rather than unified and robust discovery platforms.

Over the past decade, remarkable progress has been achieved in advancing both GNNs and SR methodologies within the realm of computational materials science. For GNNs, a diverse array of architectures has emerged, ranging from message passing neural networks (MPNN)\cite{bib18} and crystal graph convolutional neural networks (CGCNN),\cite{bib34} to atomistic line graph neural networks (ALIGNN)\cite{bib2} and graphormer,\cite{bib25,bib10,bib11} each pushing the boundaries of accuracy in predicting material properties. Concurrently, substantial efforts have been dedicated to refining GNN components, including feature engineering,\cite{bib5,bib7,bib8,bib9} algorithms for updating node and edge attributes,\cite{bib23,bib29,bib14} and readout functions,\cite{bib15} all aimed at enhancing the alignment of graph based representations with fundamental physical principles. On the other hand, SR has found significant success in deriving interpretable mathematical descriptors for specific material systems, such as perovskites, where it has guided the design of materials with optimal band gaps for photovoltaics or enhanced catalytic activity for oxygen evolution reactions.\cite{bib1,bib3} Despite these advancements, a critical gap remains: the lack of integrated framework that combines GNNs and SR to simultaneously achieve model enerality, high predictive accuracy, and interpretable descriptor generation. Current approaches in material screening and guided synthesis often prioritize one aspect at the expense of others, highlighting the need for a unified methodology that harmonizes the strengths of both paradigms to accelerate materials discovery and design.

In this work, we present a computational framework that synergistically integrates graph neural networks (GNNs) and symbolic regression (SR) facilitating advancements in the understanding of structure property relationships in materials science. Our approach introduces a sophisticated feature encoding algorithm within the GNN architecture, wherein each physical quantity feature is assigned independent weight parameters and subsequently aggregated in a high-dimensional latent space. This unique self-adaptable Graph Attention Networks combined with SR (SA-GAT-SR) formulation enables the generation of importance coefficients (ICs) at both atomic and crystallographic levels, facilitating rigorous feature screening and offering unprecedented physical interpretability of deep learning models. The re-weighted feature representations, augmented by the predictive outputs from our GNN module, are then processed through a highly optimized SR framework to extract explicit mathematical descriptors. The analysis of the resulting symbolic expressions derived from our framework, elucidates the fundamental mathematical relationships governing material property variations. This integrated paradigm effectively bridges the gap between the predictive accuracy of deep learning approaches and the interpretability of traditional scientific modeling techniques, thereby offering a robust platform for accelerated materials discovery and design.

\section*{Results and discussion}
\subsection*{Joint paradigm of graph attention network and symbolic regression}

\begin{figure}[H]
    \centering
    \includegraphics[width=1\textwidth]{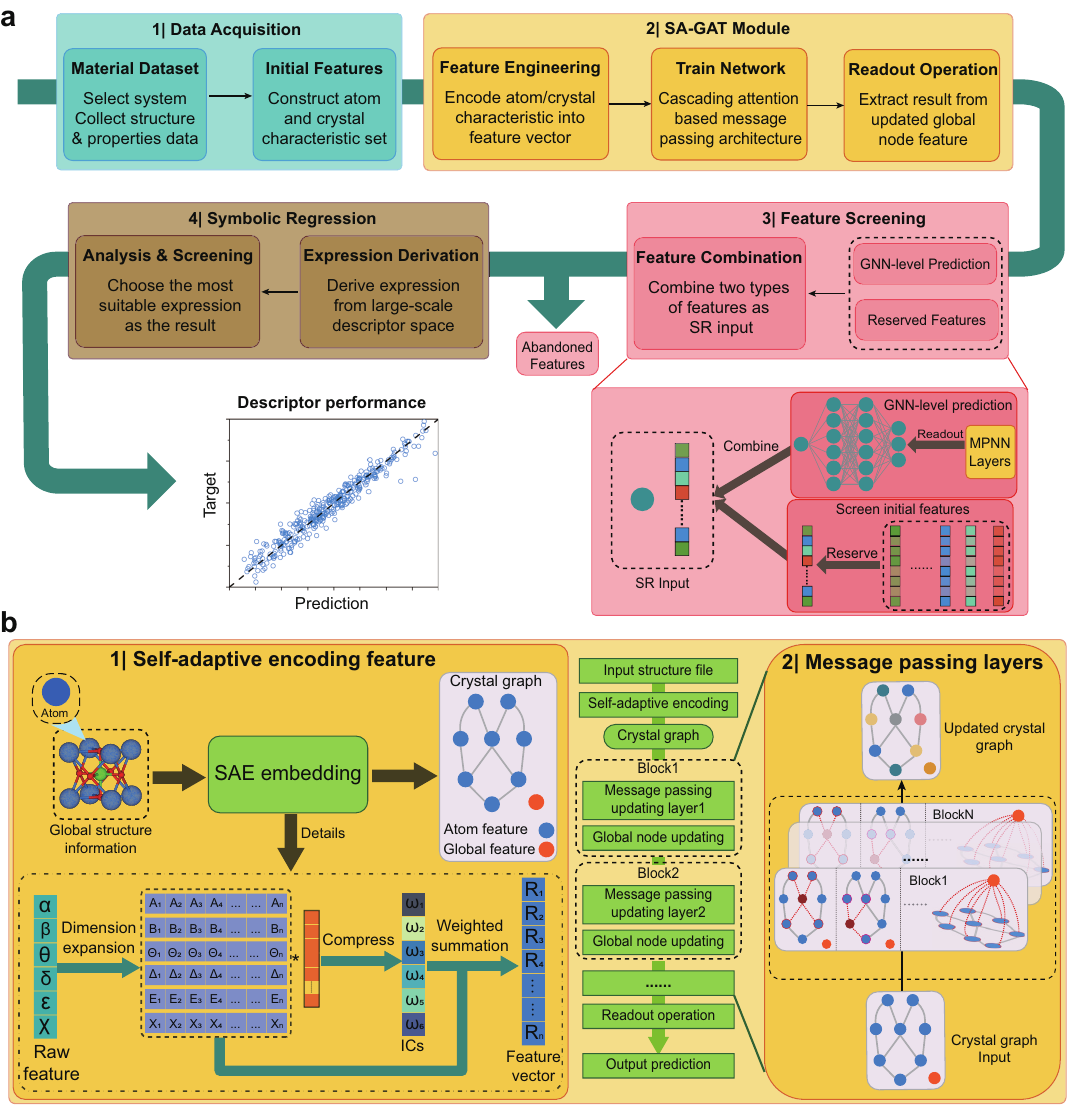}
    \caption{
        \textbf{The SA-GAT-SR model for materials prediction.} \textbf{a} The flowchart of SA-GAT-SR model encompasses four sequential stages: data acquisition (blue), preliminary GNN prediction (yellow), feature screening (pink), and symbolic regression (brown). The combination of GNN prediction and reserved features in the feature screening step is used as the input features of SR module. \textbf{b} The architecture of the GNN module. In the self-adaptable encoding (SAE) algorithm, the raw feature vector consists of scalar properties associated with atoms and unit cells from the crystal structure. The SAE assigns a weight to each characteristic and generates the initial feature vector. The blue and orange circles represent atomic and global node features, respectively. The message-passing layers include stacked node update modules, as illustrated, allowing iterative updating of feature vectors through the GNN architecture.
    }\label{Fig1}
\end{figure}

The SA-GAT-SR model is an end-to-end framework that takes crystal structures and associated features as input. Compared to previous works, which are predominately based on deep learning (DL) methods, and aim to develop universal models capable of predicting material properties across diverse systems\cite{bib6,bib32,bib33,bib16}, the proposed SA-GAT-SR methodology provides both a mathematical expression and predictive results that offer physical insight. Additionally, the model outputs an importance ranking of the initial features. Figure \ref{Fig1}a illustrates the flowchart for training a joint prediction model for specific material systems using the SA-GAT-SR approach, which can be broadly divided into four key steps. The data acquisition stage involves gathering materials from the system of interest to form the high quality training dataset. To construct node feature vectors, the dataset also requires a large set of corresponding atoms and crystal characteristics including those that construct the best descriptors as much as possible. Subsequently, the feature engineering algorithm respectively converts the material structure information and physical characteristics into a graph representation and nodes, global features and produce the ICs of each characteristics. After that, we obtain the well-trained GNN model and the corresponding GNN-level prediction result, which completes SA-GAT step. According to the ICs and initial prediction results given by pre-stage GNN model, the third level filters atomic and global features based on the specified number of reserved features. The input features of SR module are combined with the reserved features and the GNN module prediction. Finally, in the fourth step, SR module derives a series of expressions based on the recombined features.

From the perspective of model architecture, the SA-GAT-SR framework consists primarily of two modules, i.e., the GNN module enhanced by self-adaptable encoding (SAE), and the SR solver module. The GNN module comprises an SAE layer, multiple message-passing and global feature updating layers, and a readout layer. The SAE layer is responsible for performing feature engineering, constructing the initial node, edge, and global feature vectors from atomic, bonding, and crystal cell features, respectively, based on ICs. The message-passing layers operate on the crystal graph, iteratively updating each feature vector. After processing through all updating layers, the readout layer produces the final output. The output from the GNN module serves as an initial approximation of the final prediction and is fed into the SR module, implemented by the sure independence screening and specifying operator (SISSO) method\cite{bib37,bib38}.

\subsection*{Self-adaptable encoding and screening of initial features}

To convert crystal structures into graph representations compatible with GNN, we encode atomic properties as feature vectors, while the connections between atoms correspond to node and edge features, respectively. The proposed one-hot encoding and linear embedding are common feature of engineering approaches, which often lead to indistinguishability between features and an over-smoothing problem. When predicting target properties across different materials, the sensitivity of the output to input features can vary significantly. Our extensive experiments reveal that feature construction has a far greater impact on predictive performance than the specific algorithms used for feature updating. Therefore, we incorporate a self-attention-based importance-weighting module within the feature engineering process (see Supplementary Note 1).

The initial feature sets, based on physical quantities, are represented as $R_{a}$ for the atomic level and $R_{c}$ for the crystal level variables respectively. Importantly, feature sets can also be constructed for distinct atomic types or other structural units. As shown in the Figure \ref{Fig1}b, the raw feature vector of an atom from the octahedra of perovskite materials is $R_{a}=\{\alpha, \beta, \theta, \delta, \varepsilon, \chi\}$, where each element corresponds to a distinct physical characteristic. To capture more complex nonlinear relationships among these characteristics, we define a projection function  $\sigma(\cdot)$ that maps each scalar feature into a high-dimensional hidden vector. This can be represented as $h_{r_i}=\sigma (r_{i}), h_{r_i}\in \mathbb{R}^{n}$, where $r_{i}$ denotes an individual characteristic in $R_{a}$. The raw hidden feature $R_{a}$ that is a vector of length 180, can be represented as $H_a=\sigma (R_a)=\{h_{r_1}, ..., h_{r_N}\}$ where N denotes the size of the initial feature set. Each hidden feature vector in $H_a$ represents a unique high-dimensional embedding of its corresponding physical characteristic. Subsequently, to evaluate the importance of each high-dimensional feature, we introduce a learnable matrix $W_{a}$. The ICs are computed as the inner product between the weight vector $W_{a}$ and each $h_{r_i}$, followed by softmax normalization to yield the attention coefficients, which are then used as weights for the hidden feature vectors. To capture the relevance of each feature to the target property, attention weights are assigned and summed to generate the initial node feature vectors in the graph representation. The feature engineering process of node features can be described by:
\begin{align}
     & {h^0} = \sum\limits_{i = 0}^N {a_i } {h_{r_i}}
    \nonumber                                         \\
     & {a_i} = softmax (f({W_a},h_{r_i}))
    \label{eq1}
\end{align}
where the $f(\cdot )$ function is used to compute the scale attention of $h_{r_i}$, and the $h^0$ denotes the initial feature vector at layer 0. The node features are calculated by SAE algorithm. The outcome directly depends on the selection of corresponding initial feature set $R_{a}$. However, the calculation method for the global feature differs slightly from that of node features. To obtain the ICs of features from different initial feature sets, we concatenate $R_a$ and $R_c$ to form the initial global feature set. By doing so, we can compute the ICs for each relevant characteristic and the global node feature vector, as outlined in eq \ref{eq1}. This approach enables the comparison of the importance between features derived from different feature sets. For various materials, the global node features integrate both atomic-level characteristics and structural material features, ensuring that the crystal graph is uniquely defined. This integration helps mitigate the risk of over-smoothing during model training, which is a common issue arising when node information is excessively aggregated.

As illustrated in Figure \ref{Fig1}, we can select highly relevant features from the initial feature set using the weight set and preliminary predictions generated by the trained GNN module. Assuming that there are $M$ initial feature sets, each associated with key atoms and structural units, they are denoted as $\varPhi = \{ R_a^1, \ldots, R_a^M \}$, along with a unique initial crystal feature set $R_c$. The size of each initial feature set is not required to be uniform, thereby allowing for the flexible selection of various physical quantities to construct these feature sets. The combined initial global feature set is represented as $\varPsi = \{ R_a^1, \ldots, R_a^M, R_c \}$. Subsequently, the ICs for all characteristics can be derived through the SAE process applied to the set $\varPsi$. The ICs set of characteristics in different initial feature sets is defined as $A=\{a_{1}, ..., a_{N}\}$ serving as the foundation for subsequent feature screening, where $N$ stands for the number of all characteristics. Based on the ICs sets $A$, we determine the importance ranking of each feature and select the top-$k$ features from different feature set, which serves as the input for the subsequent SR module. The number of retained features per set can be independently adjusted according to specific requirements. In practice, reducing the number of selected features significantly accelerates the SR process highlighting critical importance of features screening of GNN module.

The input feature set for the SR module is defined as $\varOmega =\{S_a^1, ..., S_a^M, g_s\}$, where $S_a^i$ and $g_s$ represent the selected features from $R_a^i$ and $g$, respectively. In this work, the SR process is implemented using the SISSO machine-learning method. Unlike other approaches that rely on Density Functional Theory (DFT) results, which have limited accuracy but fast calculation speed and still demand substantial computational resources, our method uses the GNN predictions as input features to the SR module. These predictions are efficiently generated by the GNN model, thereby relaxing the strict requirement for material datasets to have DFT-based pre-calculated results. In addition to the number of screened features, the dimensionality and complexity of the descriptors significantly influence both the accuracy of the predictions and the computational efficiency of the SR process (see Supplementary Fig. 1). Careful optimization of these parameters is essential to achieving a balance between performance and resource consumption.

\subsection*{Application of the Model in Single Oxide and Halogen Perovskite Materials}

In the following computational experiments, we concentrate on predicting multiple properties of single-oxide and single-halogen perovskites with the general formulae $\mathrm{ABO}_3$ and $\mathrm{ABX}_3$ (X=F, Cl, Br, I) using the JARVIS-DFT\cite{bib35} dataset (version 2021.8.18). This dataset comprises extensive materials and provides a comprehensive set of solid-state properties ideally suited for model training and evaluation. In total, we curated a subset of 689 perovskite materials for our studies. The hyperparameter configurations of the GNN and SR modules in the SA-GAT-SR model are summarized in Supplementary Table 1. The SR module takes a total of 15 input features, which include the prediction output from the GNN module. To simplify the final expressions and accelerate the SR derivation speed, we set the unified descriptors complexity and expression dimension to 1 and 3 for all prediction tasks, respectively. The train-validation-test split ratio was consistently maintained at 85\%:10\%:5\% level. From the extensive set of properties available in this dataset, we select the following key attributes for evaluation: bandgaps corrected by the optimized Becke88 functional with van der Waals interaction ($\text{OptB88vdW}$)\cite{bib39}, formation energies, dielectric constants without ionic contributions ($\epsilon_x$, $\epsilon_y$, $\epsilon_z$), Voigt bulk modulus (Kv) and shear modulus (Gv), total energies, and energies above the convex hull ($E_{hull}$).

For the initial feature set, we select 7 physical quantities for each element and 9 physical quantities for crystal, resulting in a combined feature vector comprising 30 features for each crystal\cite{bib4,bib19}. However, since the X site of all $\mathrm{ABO}_3$ materials is fixed as oxygen, we focus on screening features from the A site and B site atoms as inputs to the SR module and the the number of initial features is 23. The symbols used in the SA-GAT-SR model are summarized in Table~\ref{tabs} and their corresponding detailed information is provided in Supplementary Table 2. The octahedral factor $G_{\mu}$ and tolerance factor $G_{t}$ are defined as $B_{ir}/r_{X}$ and $\frac{{{A_{ir}} + {r_X}}}{{\sqrt 2 ({B_{ir}} + {r_X})}}$, respectively, which are commonly used for describing perovskite structures, where the ${r_X}$ denotes the ionic radius of oxygen or halogen anion. Both features can, to a certain extent, reflect the stability of the entire crystal structure. Moreover, the ratio of these two features has been shown to exhibit a strong correlation with several functional properties of perovskite materials\cite{bib3}. As a result, this ratio is an initial feature in our analysis. Multiple studies have demonstrated that lattice structural distortions in perovskite materials can significantly influence a range of their properties\cite{bib36}. To characterize these distortion characteristics, we select a representative $BX_6$ octahedron and compute the mean B-X-B bond angle between its six neighboring octahedra, denoted as $\varTheta_{BXB}$. To further enhance the stability and interpretability of the calculated values, we adopt the cosine of $\varTheta_{BXB}$ as a distortion feature. This cosine-based measure is preferred as it maps the bond angle distortion to a range between -1 and 1, providing a naturally normalized representation of the structural distortion. Further analysis of these distortion features is provided in Supplementary Figs. 2 and 3.

\begin{table}[h]
    \caption{\textbf{The SA-GAT-SR model initial features for predicting $\mathrm{A}\mathrm{B}\mathrm{O}_3$ materials containing A site, B site, X site and global characteristics.}}\label{tabs}
    \begin{tabular*}{\textwidth}{@{\extracolsep\fill}lcccccccc}
        \toprule%
        & \multicolumn{2}{@{}c@{}}{A site} & \multicolumn{2}{@{}c@{}}{B site} & \multicolumn{2}{@{}c@{}}{X site} & \multicolumn{2}{@{}c@{}}{Global} \\
        \cmidrule(r){2-3}\cmidrule(r){4-5}\cmidrule(r){6-7}\cmidrule(r){8-9}%
        Meaning & Symbol & Unit\footnotemark[1] & Symbol & Unit\footnotemark[1] & Symbol & Unit\footnotemark[1] & Symbol & Unit\footnotemark[1] \\
        \midrule
        pauli electronegativity & $A_{en}$ & eV & $B_{en}$ & eV & $X_{en}$ & eV & - & -\\
        electron affinity & $A_{ea}$ & eV & $B_{ea}$ & eV & $X_{ea}$ & eV & - & -\\
        first ionization energy & $A_{ie1}$ & eV & $B_{ie1}$ & eV & $X_{ie1}$ & eV & - & -\\
        atomic mass & $A_{am}$ & amu & $B_{am}$ & amu & $X_{am}$ & amu & - & -\\
        atomic radius & $A_{ar}$ & $\text{Å}$ & $B_{ar}$ & $\text{Å}$ & $X_{ar}$ & $\text{Å}$ & - & -\\
        average ionic radius & $A_{ir}$ & $\text{Å}$ & $B_{ir}$ & $\text{Å}$ & $X_{ir}$ & $\text{Å}$ & - & -\\
        oxidation numbers & $A_{Q}$ & - & $B_{Q}$ & - & $X_{Q}$ & - & - & -\\
        crystal density & - & - & - & - & - & - & $G_{d}$ & $g/cm^3$\\
        crystal volume & - & - & - & - & - & - & $G_{v}$ & $\text{Å}^3$\\
        lattice vector\footnotemark[2] & - & - & - & - & - & - & $G_{l}$ & $\text{Å}$\\
        tolerance factor  & - & - & - & - & - & - & $G_{t}$ & -\\
        octahedral factor & - & - & - & - & - & - & $G_{\mu}$ & -\\
        ratio factor & - & - & - & - & - & - & $G_{\mu/t}$ & -\\
        lattice distortion  & - & - & - & - & - & - & $G_{dis}$ & -\\
        \botrule
    \end{tabular*}
    \footnotetext[1]{"-" in the Unit column denotes that the feature is a dimensionless characteristic.}
    \footnotetext[2]{The lattice vector represents 3 features containing the lengths of the lattice, i.e. (a, b, c).}
\end{table}

The performance of the SA-GAT-SR model on the $\mathrm{ABO}_3$ and $\mathrm{ABX}_3$ datasets is summarized in Supplementary Tables 3 and 4, respectively. We employ the mean absolute error (MAE) metric to evaluate regression accuracy. Furthermore, to validate the robustness of the model, we combine both datasets and assesse the model's performance on the merged dataset, with the results presented in Table~\ref{tabresOX}. The merged dataset can be denoted as $\mathrm{AB(O|X)}_3$. For comparison, we also train and test other state-of-the-art models, including Graph Attention Network (GAT)\cite{bib40}, CGCNN, and ALIGNN, using the same dataset and identical train-validation-test splits on each dataset (see Supplementary Note 2). The SA-GAT-SR model integrates a GNN module with an SR module, where the final prediction accuracy is largely depends on the performance of the GNN component. To further assess the contribution of each module, we conducted ablation study to separately evaluate their individual performances. For the SR module, the input feature set remains consistent with that of the SA-GAT-SR model, except for the initial prediction results from the preceding GNN stage. In most cases, the SA-GAT-SR model demonstrates superior performance compared to the control group models. The lowest MAE for bandgap ($E_{gap}$) with SA-GAT-SR is 0.147 eV, which outperforms GAT, CGCNN and ALIGNN by 19.7$\%$, 18.3$\%$ and 10.9$\%$, respectively. Furthermore, the full SA-GAT-SR model builds upon this foundation, achieving additional improvements in accuracy. Without the predictions from the GNN module, the MAE for $E_{gap}$ of individual SR module struggles to achieve 0.768 eV accuracy. However, in certain cases, the SR module outperforms the GNN module, which is shown in Supplementary Table 4. The MAE for $E_{hull}$ using only the SR module is 0.075 eV outperforming the 0.090 eV achieved by the GNN module alone. To assess space complexity, we analyze the number of trainable parameters in each model. The SA-GAT-SR model contains a total of 6.303M parameters, primarily due to the feature embedding matrices, which scale with the size of the input feature set and are not involved in the model's training speed. In contrast, ALIGNN comprises only 4.027M parameters. Notably, the trainable core of SA-GAT-SR includes just 3.351M parameters. Moreover, when the hidden feature dimension is set to 128, the parameter count of the core model can be reduced to as few as 1.011M without a significant loss in prediction accuracy.

\begin{table}[h]
    \caption{\textbf{Regression model performances on the $\mathrm{AB(O|X)}_3$ dataset for 9 properties using GAT, CGCNN, ALIGNN and SA-GAT-SR models.}}\label{tabresOX}
    \begin{tabular*}{\textwidth}{@{\extracolsep\fill}lcccccccc}
        \toprule%
        & \multicolumn{5}{@{}c@{}}{} &\multicolumn{3}{@{}c@{}}{Ours}\\\cmidrule{7-9}%
        Property & \#Mater. & Units & GAT & CGCNN & ALIGNN & Only GNN & Only SR & SA-GAT-SR\\
        \midrule
        Formation Energy & 689 & eV/at. & 0.515 & 0.240 & 0.105 & 0.109 & 0.318 & \textbf{0.101}\\
        Bandgap (OPT) & 689 & eV & 0.183  & 0.180 & 0.157 & 0.194 & 0.768 & \textbf{0.147}\\
        Total Energy & 689 & eV/at. & 0.280  & 0.145  & 0.123 & 0.126 & 0.644 & \textbf{0.115}\\
        $E_{hull}$ & 689 & eV & 0.129  & 0.078  & 0.071 & 0.064 & 0.161 & \textbf{0.044}\\
        Bulk Modulus Kv  & 362 & GPa & 20.18  & 22.63  & 10.37 & 10.20 & 17.99 & \textbf{9.28}\\
        Shear Modulus Gv  & 362 & GPa & 20.11  & 17.74  & 9.85 & 10.65 & 17.53 & \textbf{6.00}\\
        $\epsilon_x$  & 576 & No unit & 14.61  & 12.34  & 12.68 & 10.10 & 21.77 & \textbf{9.47}\\
        $\epsilon_y$  & 576 & No unit & 14.49  & 11.31  & 12.09 & 11.49 & 21.39 & \textbf{9.39}\\
        $\epsilon_z$  & 576 & No unit & 14.56  & 12.70  & 12.40 & 10.65 & 22.69 & \textbf{9.27}\\
        \midrule
        \#Param. & - & - & 74.43K & 0.103M & 4.027M & 6.303M &-&-\\
        \botrule
    \end{tabular*}
\end{table}

\begin{figure}[H]
    \centering
    \includegraphics[width=0.8\textwidth]{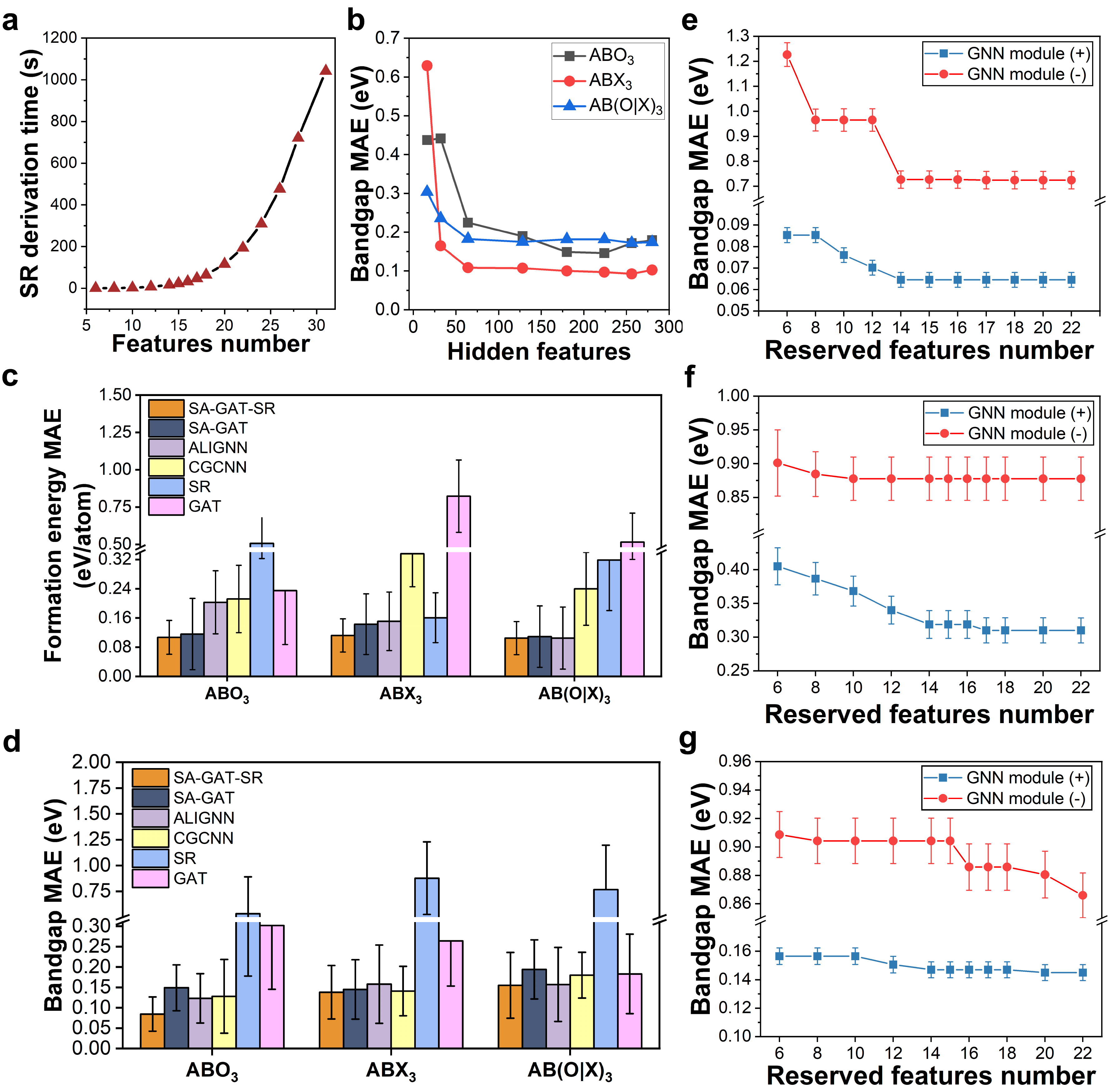}
    \caption{
        \textbf{The performance of the SA-GAT-SR model across three datasets.} \textbf{a} Dependence of the SR derivation time on the number of input features. While keeping the dimensionality and complexity of the final expressions constant, the derivation time shows exponential growth as the number of input features increases. \textbf{b} The bandgap MAE performance of the SA-GAT model with different hidden feature dimensions across three datasets. \textbf{c-d} The performance of all methods—including SA-GAT-SR, SA-GAT, ALIGNN, CGCNN, SR, and GAT—on bandgap and formation energy prediction is evaluated. SA-GAT corresponds to the GNN module within SA-GAT-SR, while SR uses the same feature set as SA-GAT-SR but excludes the GNN-predicted output. \textbf{e-g} Bandgap prediction performance, comparing the SA-GAT-SR model with and without the GNN module across three datasets as the number of reserved features increases. \textbf{e}, \textbf{f}, and \textbf{g} present the results of our study on the $\mathrm{ABO}_3$, $\mathrm{ABX}_3$ and $\mathrm{AB(O|X)}_3$ datasets, respectively.
    }\label{Fig2}
\end{figure}

To evaluate the impact of the GNN module within the SA-GAT-SR framework on the final results, we conducted another ablation study, with the findings summarized in Figure \ref{Fig2}. This experiment primarily investigates the influence of the number of selected features on the SR module's derivation time and the prediction MAE for formation energy. For the convenience of expression, the reserved features combination of bandgap with the number of $n$ is denoted as $\text{BG}\{n\}$, where the total number of the input features of SR module is actually $n+1$ including the prediction of GNN model. Figure \ref{Fig2}a demonstrates that the derivation time increases sharply as the total number of features grows (Supplementary Note 3 and Supplementary Table 5). Notably, when the feature set size reaches 22, the SR derivation time exhibits exponential growth, rendering it impractical to compute descriptors for larger datasets. It should be emphasized that the exponential growth depends solely on the number of features and is independent of the specific types of features selected, making the SAE feature selection method suitable for scenarios involving varying feature sets. In the feature screening process of the SAE module, 16 features are retained as input for the SR module to enhance accuracy, resulting in a derivation time of 32.59 seconds. In contrast, the conventional SR process requires retaining most features to ensure generalization, such as using 28 features, which takes 720.95 seconds. This demonstrates that the SAE algorithm improves the efficiency of SR by a factor of 23 compared to the conventional SR process while simultaneously enhancing accuracy.

\begin{figure}[H]
    \centering
    \includegraphics[width=1\textwidth]{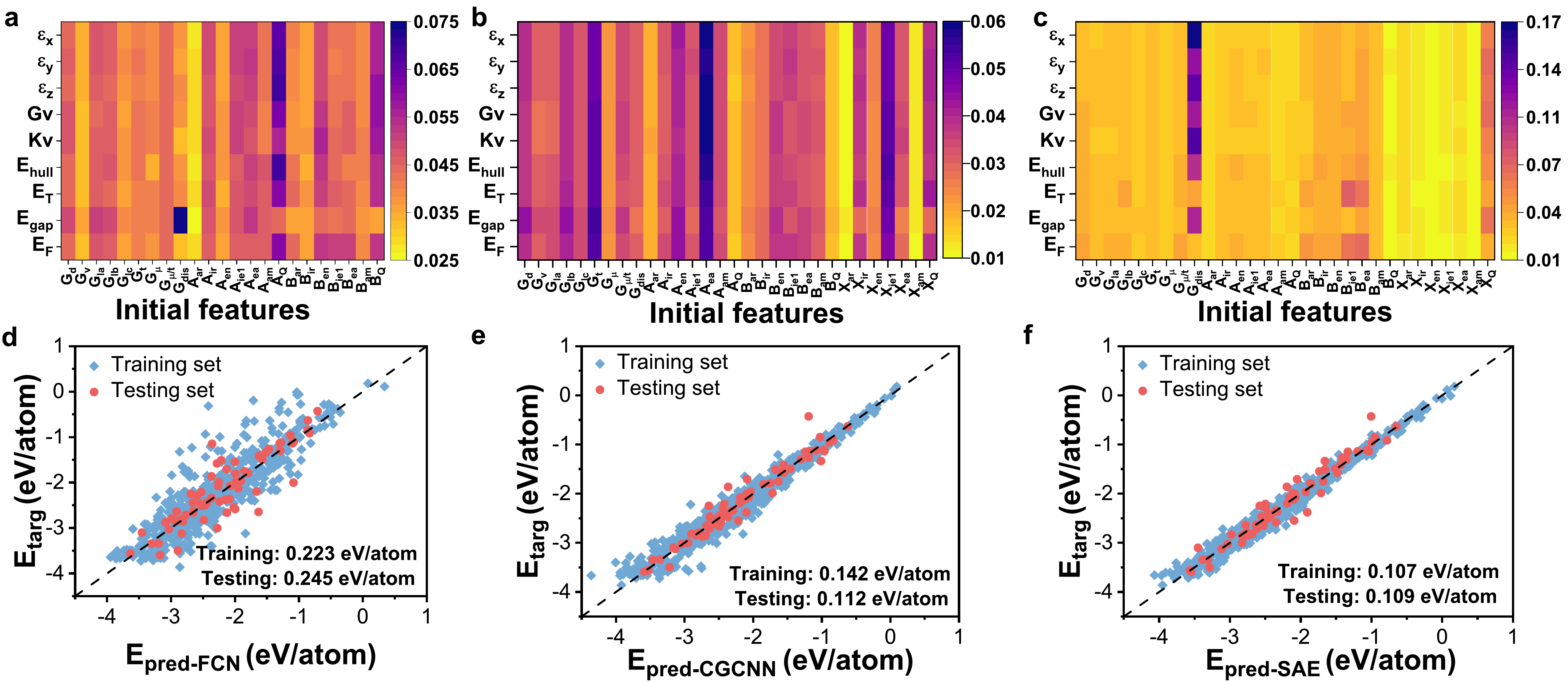}
    \caption{
        \textbf{The impact of the GNN module in both ICs Computation and formation energy prediction.} \textbf{a-c} The ICs derived by the SAE algorithm within the GNN module, where \textbf{a}, \textbf{b}, and \textbf{c} correspond to the ICs results for $\mathrm{ABO}_3$, $\mathrm{ABX}_3$, and $\mathrm{AB(O|X)}_3$, respectively. For each property prediction task, the IC reflects the significance of a specific feature within its feature set. In the figures, $E_{F}$, $E_{gap}$, and $E_{T}$ denote formation energy, bandgap, and total energy, respectively, for convenience. \textbf{d-f} Comparison of the GNN model with different three feature embedding algorithms in formation energy prediction on the $\mathrm{AB(O|X)}_3$ dataset. The blue diamonds and red dots represent the results on the training set and testing set, respectively. The GNN model with the fully connected network (FCN) embedding algorithm serves as a baseline, reflecting standard performance. The model with the CGCNN embedding algorithm utilizes one-hot encoding followed by the FCN to generate feature vectors.
    }\label{Fig3}
\end{figure}

To demonstrate the influence of the hidden feature dimension on the performance of SA-GAT model, we conduct a study on the bandgap prediction across three datasets. The results are shown in Figure \ref{Fig2}b. A dimension that is too small can in under-fitting, whereas a dimension that is too large yields only marginal improvements in performance. As the dimension comes to 128, the MAE stabilizes around 0.190 eV, 0.107 eV and 0.175 eV across three datasets. This indicates that each feature vector of node and edge is sufficiently expressive to capture the characteristics of real atoms and bonds. Additionally, in order to find out the adventage of SA-GAT-SR model against the traditional SR method, we conduct another ablation study. As shown in Figure \ref{Fig2}c and d, the GNN initial approximation method of SA-GAT-SR model has much lower MAE and higher stability compared to traditional method. Withing the SA-GAT-SR framework, the SR process serves as a second-level approximation building upon the prediction results of the GNN module and delivering further improvements in the overall prediction accuracy. In addition to the advantage of making the results more accurate, the screening capability of SAE algorithm can filter out less relevant features and reduce the size of SR input features. As for the prediction of formation energy and bandgap in $\mathrm{AB(O|X)}_3$ dataset, the SA-GAT-SR model achieves lowest MAE of 0.101 and 0.147, respectively. The SA-GAT—GNN module of SA-GAT-SR model—also suggests comparable performance compared with CGCNN and ALIGNN achieving MAE of 0.109 and 0.194. As shown in Figure \ref{Fig2}e, f and g, as the number of reserved feature increases, the feature search space expands accordingly. Therefore, as the number of input features increases, the MAE eventually converges to a fixed value, because the input feature set becomes sufficiently large to encompass all features required for the optimal solution. The speed of MAE convergence can thus serve as a valuable metric for evaluating the quality of the selected features. SR process with the GNN module exhibits stable performance $\text{BG}14$, while simultaneously maintaining a high level of prediction accuracy. Without the GNN module, the expression result in $\mathrm{AB(O|X)}_3$ dataset remains unstable until $\text{BG}22$, while results in $\mathrm{ABO}_3$ and $\mathrm{ABX}_3$ datasets stabilize earlier but continue to exhibit high MAEs. The performance of the SA-GAT-SR prediction across all target properties is provided in Supplementary Fig. 4.
Despite the fact that the SAE mechanism does not directly enhance the accuracy of the final results, it provides valuable insights by reflecting the importance of each initial feature, thus offering physical interpretation. The ICs for each feature across all prediction tasks are shown in Figure \ref{Fig3}a-c, where the values represent the relative importance ratios. Although the $\mathrm{AB(O|X)}_3$ dataset encompasses all materials in $\mathrm{ABO}_3$ and $\mathrm{ABX}_3$, the feature distributions differ across the three datasets. Consequently, the relative importance of the same initial feature varies between datasets. Similarly, the IC distributions for different properties are distinct, reflecting the physical significance of different features. In the $\mathrm{ABO}_3$ dataset, $A_Q$ shows a relatively high contribution to all properties except for bandgap with IC varying from 0.0556 to 0.0689. In contrast, $G_{dis}$ exhibits a high IC of 0.0739 for bandgap, indicating that crystal structure distortion significantly influences this property. For the $\mathrm{AB(O|X)}_3$ dataset, $G_{dis}$ also demonstrates a high IC from 0.1029 to 0.1683 across all properties, except formation energy and total energy.

To further evaluate the role of the SAE algorithm, we replace the SAE module with a fully connected network (FCN) and the CGCNN one-hot encoding method for feature embedding. Comparative experiments are conducted on the same dataset using identical hyperparameters. The MAE values obtained using the FCN, CGCNN, and SAE algorithms are denoted as $E_{\text{pred\_FCN}}$, $E_{\text{pred\_CGCNN}}$, and $E_{\text{pred\_SAE}}$, respectively. As shown in Figure \ref{Fig3}d-f, the GNN model based on SAE significantly outperforms those using FCN and CGCNN with the best MAE of 0.107 eV/atom. Among them, the FCN encoding demonstrates the poorest performance, yielding an MAE of 0.223 eV/atom. Additionally, it exhibits an overfitting issue, as evidenced by the MAE in the testing set being 0.022 eV/atom higher than that in the training set. The CGCNN encoding method shows suboptimal MAE of 0.142 eV/atom due to the limited information used to distinguish different atoms, as it relies solely on the atomic number. This limitation results in underfitting, as indicated by the MAE in the testing set being 0.030 eV/atom lower than that in the training set and prevents it from capturing the critical importance of features. Furthermore, the performance of the final descriptors depends on the predictive accuracy of the GNN module. Consequently, the SAE method yields the most accurate results and best robustness with a minimal difference of 0.002 eV/atom between the training and testing sets.
Compared to DL based methods, the interpretability of our SA-GAT-SR model primarily lies in the final derived mathematical expressions. The expression result of the SA-GAT-SR model is composed of material-specific combination feature descriptors (e.g., $\frac{B_{en}}{A_{en}}$) and a neural network-derived descriptor ($E_{P}$). Notably, the relationship between $E_{P}$ and the initial material features is nonlinear. Changes in the initial features not only affect the final output of the expression, but also alter the value of the coupled neural network descriptor $E_{P}$. Let the feature set comprising the initially selected physical quantities excluding $E_{P}$ be denoted as $M_0$. If we define the mapping relationship between a material's structure and its properties as $\mathcal{F}(\cdot)$, the SA-GAT-SR model result can be represented as follows:
\begin{equation}
    {\hat Y} = \mathcal{F}(E_{P}(M_0),M_0)
    \label{eq11}
\end{equation}
where $E_{P}(\cdot)$ represents the function describing the relationship between the GNN module's predictions and the initial features. The above expression can be directly applied for material's feature screening tasks. However, when using the model for guiding new material development, it is crucial to understand how the predicted results are influenced by each feature. The predicted value from the GNN model cannot be directly used as a guiding descriptor because it lacks an explicit mathematical relationship with the feature set $M_0$. In most experimental outcomes, the coefficient of $E_{P}$ is typically close to 1. Consequently, the final expression can be interpreted as a sum of the prediction and correction terms.

To ensure that the derived expressions are interpretable and easy to analyze, we constrain the complexity of each descriptor to 1. In the bandgap prediction experiment, equation. \ref{eq12} represents a result that achieves both high accuracy and simplicity.
\begin{equation}
    {\hat Y_{bandgap}} = {E_p} - 0.13{G_{dis}} + 0.20{X_Q} + 0.26
    \label{eq12}
\end{equation}
The expression incorporates both $G_t$ and $G_{dis}$, which have high ICs and are factors related to the stability of crystal structure. Among the terms, the coefficient of $E_{P}$ is 0.997, which can be reasonably approximated as 1. The $G_{dis}$, which has a value range from -1 to 1, appeares with a small coefficient of -0.13 in the expression, indicating that it has only a minor impact on the predicted value. Therefore, the descriptor has a value range from -0.13 to 0.13. In contrast, $X_Q$ has relatively larger coefficient of 0.20. Given that the mean value of $X_Q$ across all materials is -1.83, the mean value of the descriptor is the same. The MAE of equation. \ref{eq12} and the corresponding GNN module is 0.159 eV and 0.194 eV, respectively, indicating that the inclusion of $X_Q$ as a key factor that have a significant impact on the improvement in predictive performance. Both parameters may have a combined influence on the bandgap variation. Based on this model, when designing perovskite materials, we can optimize the bandgap by tuning the lattice distortion ($G_{dis}$) and the tolerance factor ($X_Q$). Additional expression results with a descriptor complexity of 1 and 2 are provided in Supplementary Table 6 and 7. The work by Wang et al. also derived an expression predicting the bandgap property of photovoltaic perovskites based on SISSO.\cite{bib1} In contrast to equation. \ref{eq12}, they use the DFT calculation results based on Perdew-Burke-Ernzerhof (PBE) functional as a feature similar to the $E_p$ in equation. \ref{eq12}, which can be expressed as
\begin{equation}
    {\hat Y'_{bandgap}} = 1.28{E_{PBE}} + 9.05\frac{{{l_X}}}{{{Q_B}}} - (0.77{Q_X} + 1.28){l_B} + 2.50
    \label{eq13}
\end{equation}
where $E_{PBE}$ denotes the PBE bandgap energy. And $l_X$ and $l_B$ represent the energy levels of lowest-unoccupied molecular orbitals (LUMO) of the atoms in B site and X site, respectively. Given that the PBE functional typically underestimates bandgaps by around 50$\%$, the coefficient of $E_{PBE}$ in equation~\ref{eq13} is accordingly much greater than 1. This equation leverages a less accurate DFT result to learn a correction term that more closely approximates the true value. In comparison, our formulation uses a GNN-predicted value—offering higher accuracy—as the initial approximation and entirely removes the dependency on per-material first-principles calculations, making it highly suitable for fast, scalable property predictions over large materials databases.

Many prior DL based approaches have attempted to develop fully universal models capable of accurately predicting any property using ultra-large-scale datasets. GNN models are often highly sensitive to the training dataset, making it challenging to improve robustness without careful tuning and data selection. In many cases, the primary objective is to identify key descriptors of the target property and leverage them to guide the discovery of new materials. Compared to neural network-based descriptors derived from crystal structures, mathematical descriptors offer greater interpretability and more actionable insights. Equation. \ref{eq12} presents a combined descriptor, $- 0.13G_{dis} + 0.20X_Q$, which effectively captures the relationship between structural factors and the target property. The SISSO algorithm generates multiple expression models with comparable predictive performance from a large descriptor space. In our experiments, we set the number of retained expression models to 50. Although equation. \ref{eq12} achieves an MAE of 0.159 eV, it is not the most accurate among the generated models. The MAE of the output models ranges from 0.1563 eV to 0.1723 eV, with the top-ranked models exhibiting similar predictive performance. In certain cases, it is not always necessary to select the expression with the best performance as the final result. A more concise and interpretable expression may be preferred, such as a descriptor with a complexity of 1. When screening materials across large-scale datasets to meet specific requirements, the expression with the best predictive performance is typically applied, as achieving the highest accuracy is paramount in such scenarios. However, in the context of new material development, simplicity and interpretability of the descriptors become more important.\cite{bib3} A simpler expression allows for a clearer and more intuitive determination of the optimal composition or structural features of the new material, facilitating the material design process.

\section*{Conclusion}

In this study we integrated GNN based deep learning methods with symbolic regression, and developed the SA-GAT-SR model, which achieves both high accuracy and interpretability. Compared to previously reported approaches, we propose a novel feature encoding algorithm that enhances automating large-scale feature screening, which reduces the number of SR input features from 31 to 14. This advancement significantly reduces the computation time of SR module achieving at least 23 fold acceleration compared to conventional SR method. Our approach simplifies and optimizes the feature requirements for input data in traditional SR based methods, making the model more versatile and accessible when constructing and working with new datasets. We test performance of formulated model for materials that include single perovskite oxides and halides with the general formula $\mathrm{ABO}_3$ and $\mathrm{ABX}_3$ (X=F, Cl, Br, I) organized in JARVIS-DFT\cite{bib35} dataset. We find that in predicating various material properties, the SA-GAT-SR model outperforms both the state-of-the-art GNN models and standalone SR models. In most property predictions, our model achieves an $R^2$ of 0.93 or higher across tested systems indicating strong predictive performance and reliability. In addition to the high accuracy of the SA-GAT-SR model, which is well suited for large-scale material screening, our approach also offers strong interpretability by generationg physically meaningful descriptors, marking a significant advancement in both performance and insight. For example, we have obtained and analyzed an accurate expression mapping the material's bandgap with two structural descriptors. With our method of combining with GNN and SR, the model's learning objective can be naturally shifted from directly predicting material properties to learning combinations of meaningful descriptors along with their associated coefficients. By embedding interpretability directly into neural architectures, these models enable a more transparent decision-making process and facilitate knowledge extraction from complex datasets. This may potentially foster data fusion bridging experimental and ab initio results that may have different fidelity and origin. This takes a significant step toward bridging the gap between deep learning and scientific discovery. Looking forward, the development of self-explanatory neural models will further enable automated and interpretable insights that will drive more efficient material design and discovery.

\section*{Methods}

\subsection*{Dataset}
In this study, we employ the dataset of $\mathrm{ABO}_3$ and $\mathrm{ABX}_3$ (X=F, Cl, Br, I) type perovskite materials from the JARVIS-DFT (Joint Automated Repository for Various Integrated Simulations)\cite{bib35} database for training and evaluating our model. The JARVIS-DFT database is a widely recognized resource that provides high-quality, density functional theory (DFT) calculated properties for a diverse range of materials. For our analysis, we selected materials that conform to the perovskite crystal structure as the dataset. Due to the presence of lattice distortions, the crystal symmetry of these materials includes not only cubic but also tetragonal, hexagonal, monoclinic, and orthorhombic phases. In total, we curated a subset of 689 perovskite materials for further study. It is worth noting that some materials in the dataset may be missing certain properties due to incomplete data. Given that the JARVIS-DFT database is regularly updated, we chose the specific version of the dataset from 2021.8.18, as it is the most commonly used version in recent studies and ensures consistency with previous research.

\subsection*{Self-adaptable Graph attention networks (SA-GAT)}

Our SA-GAT-SR model is a variant architecture based on GAT\cite{bib28} for the specificity of material data, especially the crystal materials. In the following, we define the crystal graph as an undirected graph with $N$ nodes, denoted by $\mathcal{G} = (V, E, g)$, where $V$ represents the node set, $E$ represents the edge set, and $g$ denotes the global node vector containing initial global features, which provides the unique representation of the graph. The node with index $i$ is represented as $v_i$, and the feature vector of $v_i$ at layer $l$ is denoted by $h_i^l$. The corresponding edge feature in $E$, which connects nodes $v_i$ and $v_j$, is represented by $e_{ij}^l$. We consider the 12 nodes within a distance of 12 $\text{Å}$ from $v_i$ as the neighboring nodes of $v_i$ and utilize a radial basis function (RBF) to compute the edge feature vectors. Additionally, we define a ranking index matrix $\mathcal{R}_g \in \mathbb{Z}^{N \times 12}$, which stores the distance ranking of each neighboring node relative to $v_i$. Each row vector of $\mathcal{R}_g$ represents the ranking index of the neighboring nodes of $v_i$.

To start, we convert the material structure into the graph representation using the self-adaptable encoding (SAE) method. Meanwhile, SAE produces the ICs of each feature for the following screening step. Then, the node features of graph are fed into the following message passing layers and iteratively updated. The normal nodes and global node representing atoms and crystal respectively are updated by different algorithms. After passing through all updating layers, we obtain the final global node feature $g^L$, where $L$ represents the total number of layers in the GNN module. The global node has learned the characteristics of the entire material and becomes a unique representation of this material. Finally, the readout function extracts deep information from $g^L$ by a feed forward network and gives out the prediction result, which is represented as:
\begin{equation}
    \hat y = FFN({g^L})
    \label{eq10}
\end{equation}
The specific form of the output $\hat{y}$ depends on the nature of the prediction task, such as classification or regression. The $\text{FFN}(\cdot)$ function includes activation functions and regularization operations, tailored to optimize the model performance according to the task requirements.

\subsection*{Atomic nodes and global node updating method}

In the GNN module, we incorporate the distance factors of different node pairs by leveraging the ranking matrix. We adopt the encoder architecture of the Transformer\cite{bib30,bib12,bib13,bib22} to update graph node features and employ a novel set2set\cite{bib17} based algorithm to update the global node feature. As shown in Figure \ref{Fig1}b, the orange global node acts as $q^*$ vector in the set2set model to represent the state of the node set. The network is composed of multiple serially stacked blocks represented by green rectangle, each comprising a message passing layer and a set2set layer, which update the atomic and global nodes, respectively.

At the start of each layer, the ranking matrix $\mathcal{R}_g$ is embedded into a distance coefficient matrix $\mathcal{C}_g \in \mathbb{R}^{N \times 12}$, represented by $\mathcal{C}_g = \text{Emb}_r(\mathcal{R}_g)$. The message passing update algorithm is formulated as follows:
\begin{equation}
    h_i^{l + 1} = h_i^l + \sum\limits_{k \in {N_i}} {{a_{ik}}W_{val}^lz_{ik}^l}
    \label{eq2}
\end{equation}
\begin{equation}
    e_{ij}^{l + 1} = e_{ij}^l + {a_{ij}}W_{edge}^l(h_i^l \oplus h_j^l \oplus e_{ij}^l)
    \label{eq3}
\end{equation}
\begin{equation}
    {a_{ik}} = Softmax (W_{qry}^lh_i^l\cdot W_{key}^lz_{ik}^l + {c_{ik}})
    \label{eq4}
\end{equation}
\begin{equation}
    z_{ik}^l = h_k^l \oplus e_{ik}^l
    \label{eq5}
\end{equation}
where $c_{ij} \in \mathcal{C}_g$ is a scaling factor representing the distance coefficient between $v_i$ and $v_j$, and $\oplus$ denotes the vector concatenation operation. ${N_i}$ denotes the neighbors of $v_i$ in graph $\mathcal{G}$, while $W_{qry}, W_{key}, W_{val}, W_{edge}$ are learnable embedding matrices. Equations \ref{eq2} and \ref{eq3} represent the updates of the node feature $h_i$ and edge feature $e_{ij}$ connecting $v_i$ and $v_j$ in layer $l+1$.

The core of the GNN module comprises multiple stacked message-passing layers that iteratively update the node and edge features. The readout operation is performed using the global node feature, which is updated at each layer. Inspired by previous work\cite{bib31,bib27}, the global node, sometimes referred to as [VNode] in other models, is an artificially introduced node. Unlike other approaches, our model's global node, which represents the entire graph is initialized from actual crystal properties via the SAE module. Thus, it acts as a representation of the graph for the readout operation but does not participate in node updates within the graph.

To enhance information retention across layers, we employ an inter-layer set2set method with a GRU\cite{bib41} updating unit. This configuration allows the readout operation to be conducted on the final output global node from the last layer, computed as follows:
\begin{equation}
    {g^{l + 1}} = {g^l} + \varphi ({q^{l + 1}} \oplus {r^{l + 1}})
    \label{eq6}
\end{equation}
\begin{equation}
    {q^{l + 1}} = GRU({g^l})
    \label{eq7}
\end{equation}
\begin{equation}
    {r^{l + 1}} = \sum\limits_{i \in N} {e_i^{l + 1}x_i^{l + 1}}
    \label{eq8}
\end{equation}
where,
\begin{align}
    e_i^{l + 1} = Softmax ({q^{l + 1}}x_i^{l + 1})
    \nonumber \\
    x_i^{l + 1} = h_i^{l + 1} + \sum\limits_{j \in N_i^1} {a_{ij}^qW_{val}^qe_{ij}^{l + 1}}
    \nonumber \\
    a_{ij}^q = Softmax (W_{qry}^qh_i^l\cdot W_{key}^qe_{ik}^{l + 1})
    \label{eq9}
\end{align}
In the above equations, $W_{qry}^q, W_{key}^q, W_{val}^q$ are learnable matrix parameters within the readout module, and ${N_i^1}$ denotes the 1-hop neighbors of $v_i$ in graph $\mathcal{G}$. The query feature vector $q^{l+1}$ in the set2set module is derived from the global node using a GRU function that encapsulates key graph information. The $e_i^{l+1}$ is computed based on $x_i^{l+1}$, obtained from node features while accounting for the local environment around each node. The function $\varphi(\cdot)$ represents a feed forward network (FFN) that converts the combined result into the output global node feature.

\section*{acknowledgement}
This work is supported by National Natural Science Foundation of China (No.12374057), and Fundamental Research Funds for the Central Universities. The work (S.T.) at Los Alamos National Laboratory (LANL) was performed at the Center for Integrated Nanotechnologies (CINT), a U.S. Department of Energy, Office of Science user facility at LANL.

\section*{Code availability}
The code and well-trained GNN model and all configurations used in this work are available
on GitHub at \url{https://github.com/MustBeOne/SA-GAT-SR}.

\bibliography{sa-gat-sr}



\section*{Supplementary material}

# Table of CONTENTS



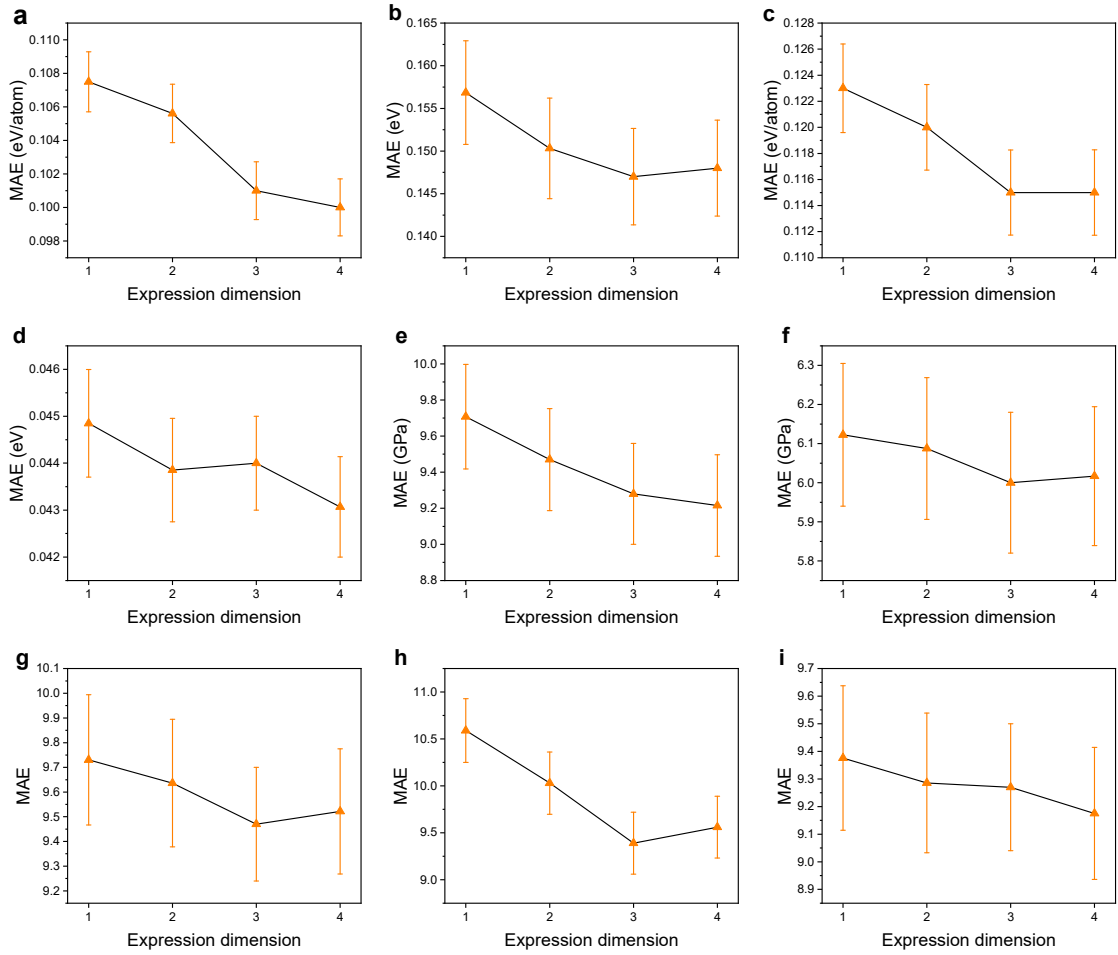

Supplementary Figure 1: **The effect of dimension of expression derived by SR module on the performance of all properties in $AB(O|X)_3$ dataset.** **a** Formation Energy. **b** Bandgap (OPT). **c** Total Energy. **d** $E_{hull}$. **e** Bulk Modulus $K_V$. **f** Shear Modulus $G_V$. **g** $\varepsilon_x$. **h** $\varepsilon_y$. **i** $\varepsilon_z$.

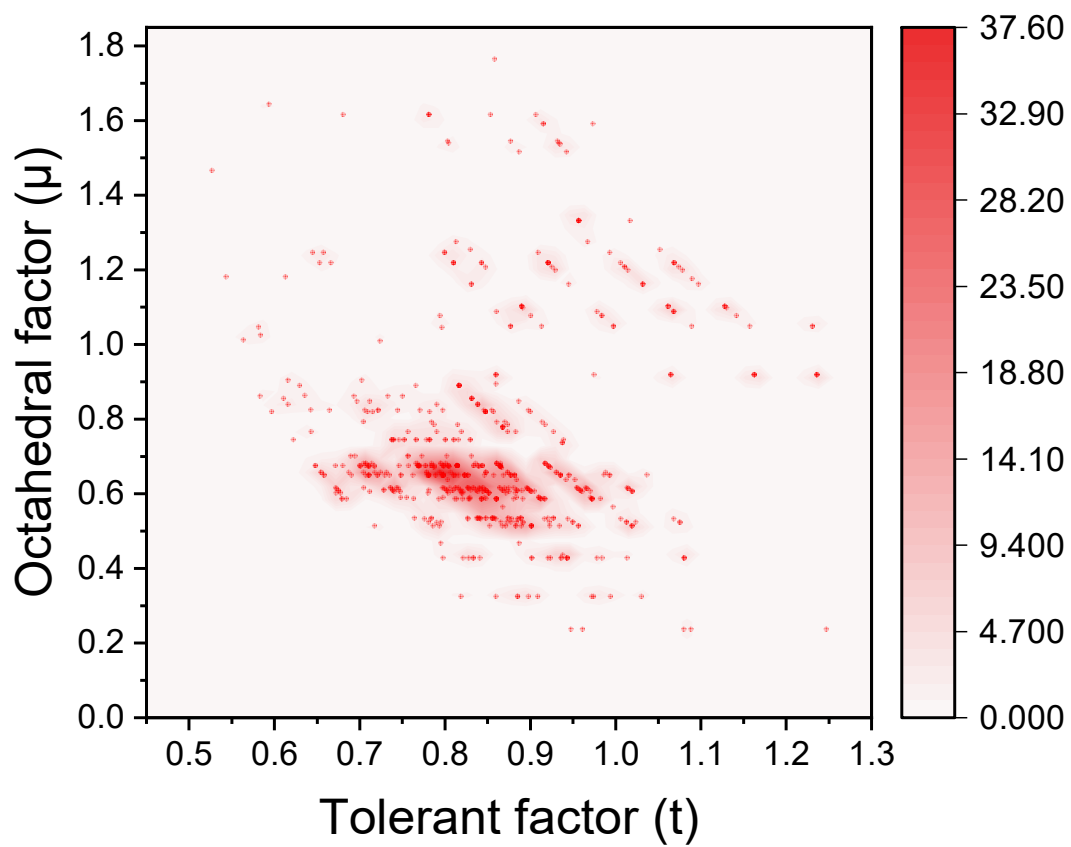

Supplementary Figure 2: **The distribution density of tolerant factor (t) and octahedral factor (µ) for each material.** Each red dot represents the position of a material in the diagram, while the red region indicates the density of the µ-t distribution.

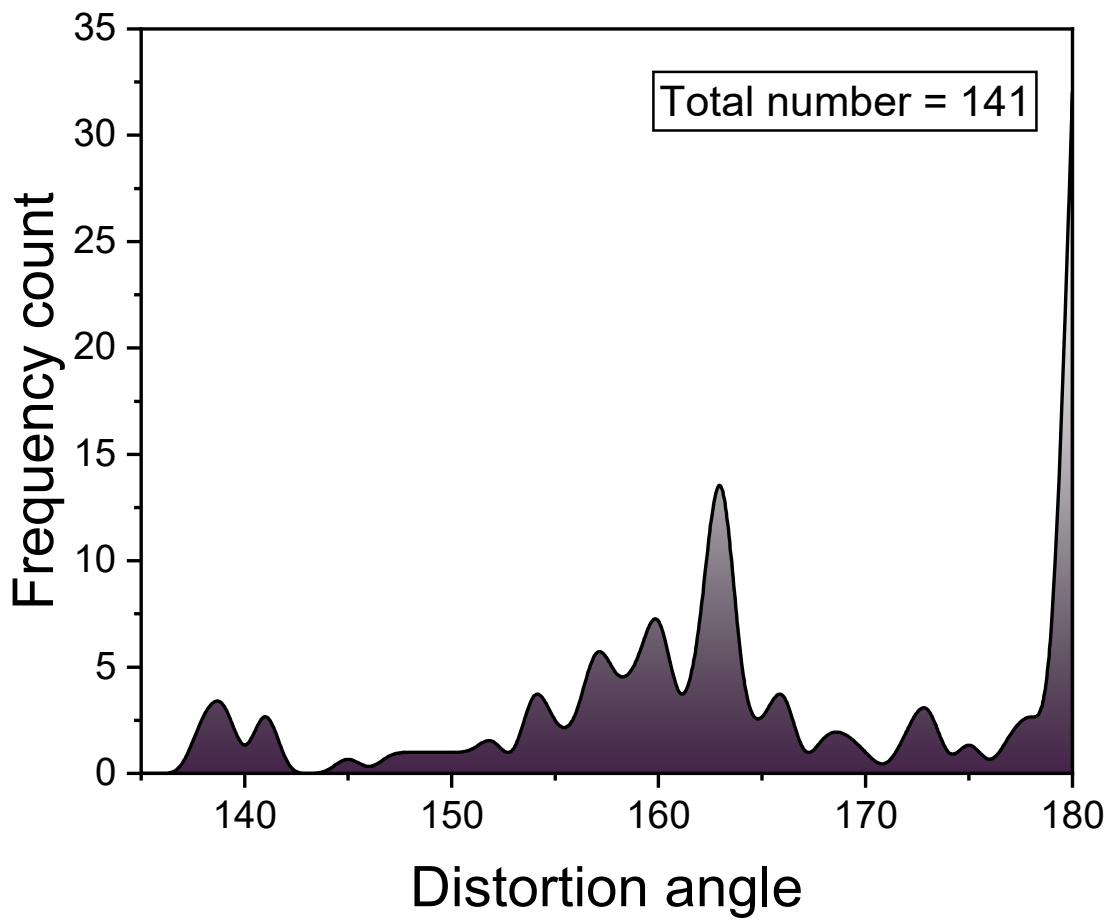

Supplementary Figure 3: **The distribution of distortion angles for materials with distortion.** A total of 141 materials exhibit distortion, while the B–X–B angle in the remaining materials is 180°. The diagram displays the frequency distribution of only the distorted materials.

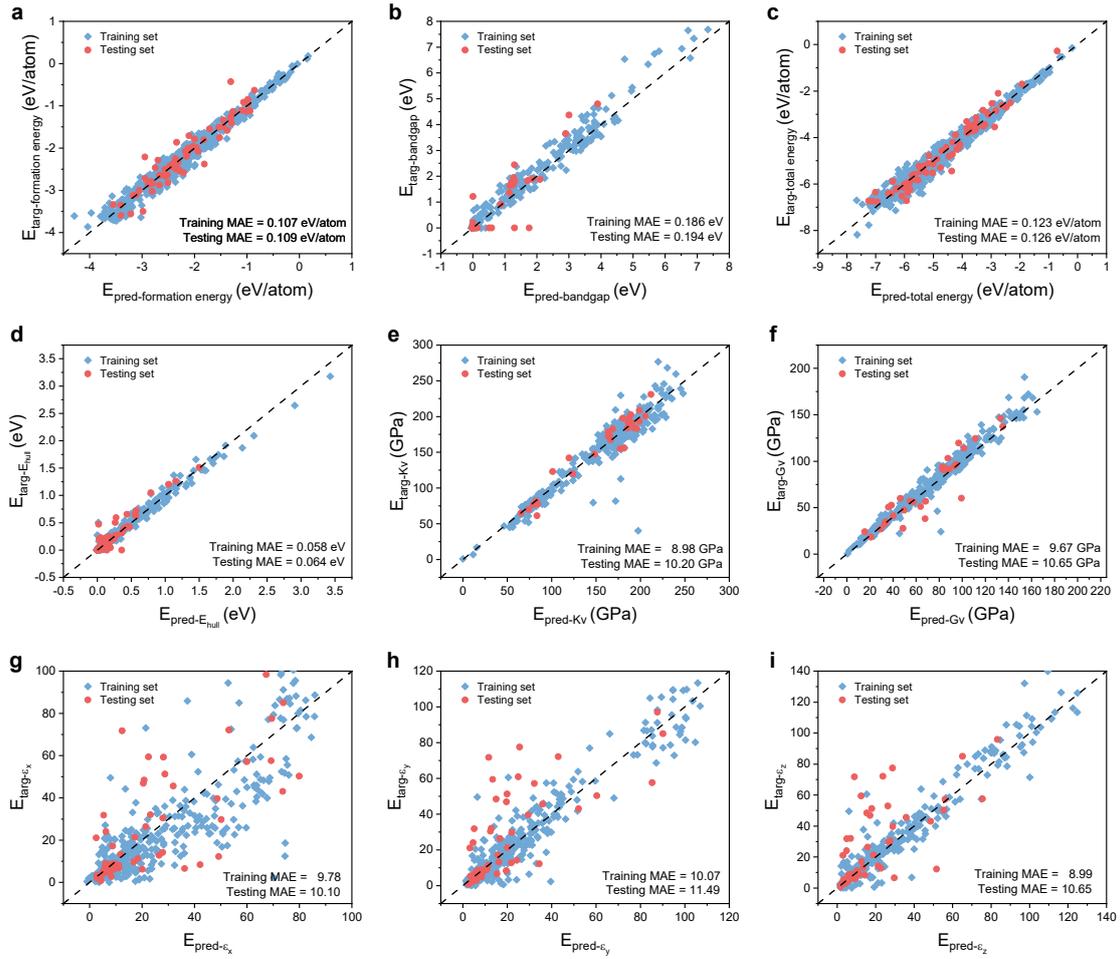

Supplementary Figure 4: **Prediction result of GNN module against target of all properties in $AB(O|X)_3$ dataset. a** Formation Energy. **b** Bandgap (OPT). **c** Total Energy. **d** $E_{hull}$. **e** Bulk Modulus $K_V$. **f** Shear Modulus $G_V$. **g** $\varepsilon_x$. **h** $\varepsilon_y$. **i** $\varepsilon_z$.

**Supplementary Table 1:** The hyperparameters configuration of SA-GAT-SR model used for the prediction of perovskite materials.

| Hyperparameter | Range |
| --- | --- |
| SAE atomic embedded features | 180 |
| SA-GAT layers | 4 |
| SA-GAT attention head | 4 |
| Hidden features [a] | 32, 64, 128, 180, 224, 256, 288 |
| Edge features | 48 |
| Learning rate | 0.00008 |
| Learning rate warmup | True |
| Optimizer | AdamW |
| Batch size | 8 |
| Maximum neighbors number | 12 |
| Dropout fraction | 0.3 |
| Reserved features number | 12 |
| Descriptors complexity | 1 |
| Expression dimension | 3 |
| Use GNN prediction | True |
| Evaluation metric [b] | RMSE |
| Operation set | '+', '-', '*', '/', '^-1' |

[a] These hidden feature dimensions are only used for experiments on their impact on the performance of SA-GAT model. We only use 180 as the feature dimension in other experiments.

[b] We use the Root Mean Square Error (RMSE) as the evaluation metric of expression results derived by SR module.

**Supplementary Table 2:** Range of properties of all materials used for the initial feature set.

| Set | Property | Unit | Range | Bin number [a] |
|---|---|---|---|---|
| Global | Crystal density $G_d$ | $g/cm^3$ | 2.59 – 12.81 | 200 |
| | Crystal volume $G_v$ | $Å^3$ | 38.10 – 472.35 | |
| | Lattice vector $G_l$ | Å | 3.36 – 8.74 | |
| | Tolerance factor $G_t$ | - | 0.53 – 1.25 | |
| | Octahedral factor $G_\mu$ | - | 0.24 – 1.77 | |
| | Ratio factor $G_{\mu/t}$ | - | 0.19 – 2.79 | |
| | Lattice distortion $G_{dis}$ | - | -1.0 – -0.19 | |
| A site | Atomic radius $A_{ar}$ | Å | 1.05 – 2.60 | 250 |
| | Average ionic radius $A_{ir}$ | Å | 0.59 – 1.81 | |
| | Pauli electronegativity $A_{en}$ | eV | 0.79 – 2.54 | |
| | First ionization energy $A_{ie1}$ | eV | 3.89 – 10.44 | |
| | Electron affinity $A_{ea}$ | eV | -0.72 – 2.31 | |
| | Atomic mass $A_{am}$ | amu | 6.94 – 244.0 | |
| | Oxidation numbers $A_Q$ | - | 1.0 – 5.0 | |
| B site | Atomic radius $B_{ar}$ | Å | 0.70 – 2.15 | 250 |
| | Average ionic radius $B_{ir}$ | Å | 0.30 – 1.49 | |
| | Pauli electronegativity $B_{en}$ | eV | 0.89 – 2.66 | |
| | First ionization energy $B_{ie1}$ | eV | 5.14 – 11.26 | |
| | Electron affinity $B_{ea}$ | eV | -0.72 – 3.06 | |
| | Atomic mass $B_{am}$ | amu | 6.94 – 244.0 | |
| | Oxidation numbers $B_Q$ | - | 1.0 – 5.0 | |
| X site | Atomic radius $X_{ar}$ | Å | 0.5 – 1.40 | 250 |
| | Average ionic radius $X_{ir}$ | Å | 0.70 – 1.27 | |
| | Pauli electronegativity $X_{en}$ | eV | 2.66 – 3.98 | |
| | First ionization energy $X_{ie1}$ | eV | 10.45 – 17.42 | |
| | Electron affinity $X_{ea}$ | eV | 1.46 – 3.40 | |
| | Atomic mass $X_{am}$ | amu | 16.00 – 126.90 | |
| | Oxidation numbers $X_Q$ | - | -2.0 -1.0 | |

[a] The number of divided bins may vary across different feature sets. A higher number of bins increases the resolution for distinguishing between different atoms and helps prevent cases where two atoms share identical feature vectors.

**Supplementary Table 3:** Performances on the $ABO_3$ dataset for 9 properties using GAT, CGCNN, ALIGNN and SA-GAT-SR models.

| Property | #MATL. | GAT | CGCNN | ALIGNN | Ours | | |
|---|---|---|---|---|---|---|---|
| | | | | | Only GNN | Only SR | SA-GAT-SR |
| Formation Energy | 570 | 0.235 | 0.212 | 0.203 | 0.116 | 0.351 | **0.108** |
| Bandgap (OPT) | 570 | 0.302 | 0.128 | 0.123 | 0.149 | 0.534 | **0.084** |
| Total Energy | 570 | 0.307 | 0.232 | 0.195 | 0.201 | 0.530 | **0.185** |
| $E_{hull}$ | 570 | 0.120 | 0.102 | 0.086 | 0.074 | 0.189 | **0.044** |
| Bulk Modulus $K_V$ | 280 | 20.27 | 19.79 | 18.76 | 17.08 | 21.10 | **11.92** |
| Shear Modulus $G_V$ | 280 | 16.38 | 10.61 | 9.90 | 12.85 | 19.01 | **5.66** |
| $\varepsilon_x$ | 472 | 23.21 | 14.54 | 14.14 | 14.38 | 23.08 | **6.87** |
| $\varepsilon_y$ | 472 | 23.39 | 15.26 | 14.56 | 14.05 | 23.06 | **6.55** |
| $\varepsilon_z$ | 472 | 24.33 | 15.91 | 15.23 | 14.35 | 22.52 | **6.98** |

**Supplementary Table 4:** Performances on the $ABX_3$ dataset for 9 properties using GAT, CGCNN, ALIGNN and SA-GAT-SR models.

| Property | #MATL. | GAT | CGCNN | ALIGNN | Ours | | |
|---|---|---|---|---|---|---|---|
| | | | | | Only GNN | Only SR | SA-GAT-SR |
| Formation Energy | 119 | 0.823 | 0.336 | 0.151 | 0.143 | 0.162 | **0.112** |
| Bandgap (OPT) | 119 | 0.264 | 0.141 | 0.158 | 0.145 | 0.878 | **0.138** |
| Total Energy | 119 | 0.259 | 0.223 | 0.190 | 0.214 | 0.466 | **0.188** |
| $E_{hull}$ | 119 | 0.169 | 0.093 | 0.062 | 0.090 | 0.075 | **0.050** |
| Bulk Modulus $K_V$ | 82 | 6.36 | 5.72 | 8.15 | 5.52 | 5.80 | **4.31** |
| Shear Modulus $G_V$ | 28 | 4.25 | 3.98 | 3.76 | 3.66 | 4.42 | **2.61** |
| $\varepsilon_x$ | 104 | 2.98 | 3.85 | 6.41 | 3.96 | 4.94 | **2.46** |
| $\varepsilon_y$ | 104 | 3.05 | 3.29 | 6.55 | 3.83 | 4.82 | **2.53** |
| $\varepsilon_z$ | 104 | 3.02 | 3.95 | 6.93 | 2.95 | 4.87 | **2.64** |

**Supplementary Table 5:** Hardware environment used in the SR module.

| Feature | Value |
|---|---|
| CPU core number | 64 |
| CPU thread number | 128 |
| CPU min frequency | 1500 (MHz) |
| CPU max frequency | 2250 (MHz) |
| Architecture | x86, 64-bit |
| Model name | AMD EPYC 7742 64-Core Processor |
| Total memory | 515728 (MB) |

**Supplementary Table 6:** The expression results derived by the SR module with descriptor complexity of 1 for all target properties in the $AB(O|X)_3$ dataset.

| Property | Derived expression |
|---|---|
| Formation Energy | $\hat{Y} = 0.97E_p + 0.016B_{ie1} + 0.028B_{ea} - 0.23$ |
| Bandgap (OPT) | $\hat{Y} = 1.06E_p + 0.20X_Q - 0.13G_{dis} + 0.26$ |
| Total Energy | $\hat{Y} = 0.94E_p + 0.15X_Q + 0.048B_{ie1} - 0.36$ |
| $E_{hull}$ | $\hat{Y} = 0.92E_p - 0.025G_{dis} + 0.013B_{ea} - 0.023$ |
| Bulk Modulus $K_V$ | $\hat{Y} = 0.88E_p - 11.68X_Q - 5.13G_{dis} - 8.79$ |
| Shear Modulus $G_V$ | $\hat{Y} = 1.00E_p + 1.61X_Q + 4.37G_{dis} + 6.78$ |
| $\varepsilon_x$ | $\hat{Y} = 0.99E_p + 2.68X_Q - 7.50G_{dis} - 4.48$ |
| $\varepsilon_y$ | $\hat{Y} = 1.24E_p + 5.02X_Q - 8.20G_{dis} - 3.44$ |
| $\varepsilon_z$ | $\hat{Y} = 1.04E_p - 14.03B_{ir} - 16.79G_{dis} - 5.05$ |

**Supplementary Table 7:** The expression results derived by the SR module with descriptor complexity of 2 for all target properties in the $AB(O|X)_3$ dataset.

| Property | Derived expression |
|---|---|
| Formation Energy | $\hat{Y} = 0.96 E_p + 0.014(B_{ea} * B_{ie1}) - 0.09 \dfrac{B_{ea}}{A_{en}} - 0.12$ |
| Bandgap (OPT) | $\hat{Y} = 0.078(E_p * B_{ie1}) - 0.30(E_p * G_{dis}) - 0.35 \dfrac{E_p}{X_Q} + 0.024$ |
| Total Energy | $\hat{Y} = 0.93(E_p - X_{en}) - 3.60 \dfrac{X_{ir}}{X_{en}} + 0.052(A_{en} + B_{ie1}) + 3.71$ |
| $E_{hull}$ | $\hat{Y} = -0.73(E_p * G_{dis}) + 0.15 \dfrac{E_p}{A_{ir}} + 0.023 \dfrac{B_{ea}}{A_{ir}} + 0.004$ |
| Bulk Modulus $K_V$ | $\hat{Y} = 0.83(E_p - B_{ea}) - 33.3 \dfrac{G_{dis}}{X_Q} + 18.8 \dfrac{B_{ir}}{G_{dis}} + 61.2$ |
| Shear Modulus $G_V$ | $\hat{Y} = 1.34(E_p - B_{en}) - 0.10 \dfrac{E_p}{B_{ir}} - 0.30(E_p * B_{ir}) + 3.66$ |
| $\varepsilon_x$ | $\hat{Y} = 0.42(E_p * A_{en}) + 0.82 \dfrac{E_p}{A_{en}} - 0.086(E_p * B_Q) - 0.94$ |
| $\varepsilon_y$ | $\hat{Y} = 1.03(E_p * B_{ir}) + 0.53 \dfrac{E_p}{B_{ir}} + 0.31 \dfrac{E_p}{G_{dis}} - 3.82$ |
| $\varepsilon_z$ | $\hat{Y} = -0.28(E_p * B_{ir}) - 1.10(E_p * G_{dis}) - 0.20 \dfrac{E_p}{G_{dis}} - 0.28$ |

**Supplementary Note 1: The method for generating feature vectors of atomic nodes and global node, along with importance factors (ICs).**

The most crucial aspect of feature vector encoding is to ensure the uniqueness of each material and its constituent atoms. A commonly used one-hot encoding method differentiates atoms based solely on their atomic numbers. However, this approach fails to effectively utilize other physical properties of atoms. In contrast, the SAE encoding method not only guarantees the uniqueness of each feature vector but also provides richer information for the subsequent node update process and final output, enhancing both representation and interpretability.

To capture the latent features of each initial scalar feature in a higher-dimensional space, we first project each scalar feature into a higher-dimensional representation. Assuming the initial feature set of a specific atom in the material is denoted as $R_a^i = \{r_a^{i,1}, ..., r_a^{i,M}\}$, this process can be expressed as:

$$H_a^i = \text{Proj}(R_a^i) \tag{1}$$

where $H_a^i = \{\varphi_a^{i,1}, ..., \varphi_a^{i,M}\}$, $R_a^{i,i} \in \mathbb{R}^{n \times 1}$. We employ the Ordinal Binning method to achieve this transformation. Suppose the lower and upper bounds of $r_a^{i,i}$ are denoted as $r_{\min}^{i,i}$ and $r_{\max}^{i,i}$, respectively. We then define the bin number, which determines the grid divisions within the interval $[r_{\min}^{i,i}, r_{\max}^{i,i}]$. This allows us to obtain the position index of $r_a^{i,i}$ within the interval, denoted as $n_{i,i}$. Finally, we apply an embedding function to map $r_a^{i,i}$ into a high-dimensional space, obtaining the hidden feature representation of $\varphi_a^{i,i}$.

Since the number of atomic species is finite, setting the bin number to a sufficiently large value ensures that each feature $r_a^{i,i}$ attains a unique position index $n_{i,i}$. Consequently, this method guarantees that every atom is assigned a distinct $H_a^i$. For the global node, its feature set can be represented as $\Psi = \{R_a^1, ..., R_a^M, R_c\}$. Given that each $R_a^i$ has a corresponding $H_a^i$, and when the bin number is sufficiently large, $R_c$ also has a unique corresponding hidden feature representation. Therefore, each material possesses a distinct hidden feature set.

After obtaining $H_a^i$, we define a learnable attention vector $\alpha_a \in \mathbb{R}^{n \times 1}$. The attention coefficient for each feature is computed by taking the inner product between $\alpha_a$ and each hidden feature vector in $H_a^i$. Subsequently, all attention coefficients are

14

normalized to obtain the ICs for each scalar feature, which can be defined as $A_a^i = \{a_a^{i,1},..., a_a^{i,M}\}$. The same operation is applied to $\Psi$. Once the model is trained, the ICs derived from the global node feature set will carry meaningful reference values, providing insights into the relative importance of each scalar feature.

The generation of feature vectors for each atomic node in the graph can be formulated as the following process:

$$h_i^0 = \sum_{i=1}^{M} a_a^{i,i} \varphi_a^{i,i} \tag{2}$$

where $h_i^0$ represents the initial feature vector of the $i$ th atom in the graph. This process can be regarded as a weighted summation of each atomic feature $H_a^i$ and its corresponding $A_a^i$. The same procedure applies to the global node. Compared to the conventional one-hot encoding method, which fails to fully utilize atomic features, this approach allows the model to leverage a broader range of atomic physical properties in its predictions. Moreover, it provides meaningful ICs, offering interpretability and reference value for feature selection.

**Supplementary Note 2: The comparative state-of-the-art models used in the experiments.**

The representative state-of-the-art models used for comparison with our SA-GAT-SR model are CGCNN and ALIGNN. Although numerous advanced models have been proposed recently, most models applied in crystal material prediction are fundamentally based on these two architectures. Furthermore, the SA-GAT-SR model introduces a novel approach that integrates characteristics from different algorithms, making direct comparisons with other models less necessary. Therefore, we select CGCNN and ALIGNN as baseline models for accuracy comparison experiments, ensuring that all models are trained with identical hyperparameters.

Regarding the GAT model, since the core algorithm of SA-GAT-SR is the self-attention mechanism, we replace the convolutional layers of CGCNN with self-attention layers, creating a GAT-based model specifically designed for crystal material prediction. Because crystal structure information cannot be directly fed into the original GAT model, we preprocess the dataset in the same way as for CGCNN and use identical hyperparameters during training. This GAT model serves as a control experiment to evaluate the impact of the attention mechanism on model performance.

To investigate the individual contributions of the GNN and SR modules in our joint model, we conduct ablation experiments by isolating each module. The standalone SR module is commonly employed in descriptor derivation studies, while the standalone GNN module represents a well-established research direction focused on developing a universal model capable of accurately predicting unseen materials. The SA-GAT-SR model experiment demonstrates the combined effect of methods from these two major fields.

**Supplementary Note 3: The hardware specifications and hyperparameters configuration used in the SR derivation time experiment.**

In the experiment on SR derivation time under different numbers of input features, the results may vary depending on the performance of the hardware used. Even with the same device, the time consumption can differ based on the control parameter settings. Therefore, it is essential to specify the experimental details and hardware environment. The relevant hardware information is provided in **Supplementary Table 6**. Detailed performance parameters can be obtained by visiting (https://www.amd.com/en/support/downloads/drivers.html/processors/epyc/epyc-7002-series/amd-epyc-7742.html).

Regarding the control parameters of the SR module, the number of processor cores allocated for running the SISSO program is set to 12 for all instances. The feature storage method in memory is configured as "**by data**" to enhance derivation speed. Additionally, the number of features in each SIS-selected subspace and the number of top-ranked models to output are fixed at 3000 and 50, respectively, for all "*****.in**" files. The parameters listed above pertain specifically to experimental performance, and further details on the usage of SISSO can be found in the instruction manual.

% 

\end{document}